\documentstyle[psfig,epsf,conf-X,art10]{article}
\def\K{{\rm\thinspace K}}

\def\Mpc{{\rm\thinspace Mpc}}
\def\Msun{\hbox{$\rm\thinspace M_{\odot}$}}

\def\h50{\hbox{$\rm\thinspace h_{50}$}}
\def\h50m1{\hbox{$\rm\thinspace h_{50}^{-1}$}}
\def\Msun{\mathop{\rm M_{\odot}}\nolimits}

\def\Mpc{\mathop{\rm Mpc}\nolimits}

\def\K{\mathop{\rm K}\nolimits}
\def\P3M{\hbox{$P^{3}M$}}

\def\AP3M{\hbox{$AdP^{3}M$}}

\def\cc2{c2}
\def\cc3{c3}
\def\cc4{c4}
\def\cc{c}

\def\apj{ApJ}
\def\apjl{ApJL}

\def\mnras{MNRAS}

\def\spose#1{\hbox to 0pt{#1\hss}}
\def\approxlt{\mathrel{\spose{\lower 3pt\hbox{$\sim$}}
	\raise 2.0pt\hbox{$<$}}}
\def\approxgt{\mathrel{\spose{\lower 3pt\hbox{$\sim$}}
	\raise 2.0pt\hbox{$>$}}}

\def\<{\thinspace}
\def\boxit#1{\vbox{\hrule\hbox{\vrule\kern3pt\vbox{\kern3pt
          #1 \kern3pt}\kern3pt\vrule}\hrule}}

\begin{document} 
\small
\heading{%
%Begin Heading
%
The effect of radiative cooling on X-ray emission from clusters of galaxies
% End Heading
}
\par\medskip\noindent
\author{%
%Begin Author names
F.~R.~Pearce$^{1}$, H.~M.~P.~Couchman$^2$, P.~A.~Thomas$^3$,
A.~C.~Edge$^{1}$
%End Author names
}
\address{%
%First address
Department of Physics, University of Durham, Durham, DH1 3LE, UK
}
\address{%
% Second Address
Department of Physics and Astronomy, McMaster University, Hamilton, Ontario, Canada, L8S 4M1
}
\address{%
% Third Address
Astronomy Centre, CPES, University of Sussex, Falmer, Brighton, BN1 9QJ, UK
}

\begin{abstract}
In this paper we use state-of-the-art N-body hydrodynamic simulations
of a cosmological volume of side $100\Mpc$ to produce many galaxy
clusters simultaneously in both the standard cold dark matter (SCDM)
cosmology and a cosmology with a positive cosmological constant
($\Lambda$CDM).  We have performed simulations of the same volume both
with and without the effects of radiative cooling, but in all cases
neglect the effects of star formation and feedback.  With radiative
cooling clusters are on average five times {\it less} luminous in
X-rays than the same cluster simulated without cooling. The importance
of the mass of the central galaxy in determining the X-ray luminosity
is stressed.
\end{abstract}
\section{Introduction}

Clusters of galaxies are the largest virialised structures in the
Universe, evolving rapidly at recent times because in hierarchical
cosmologies big objects form last. Even at moderate redshifts the
number of large dark matter halos in a cold dark matter Universe with
a significant, positive cosmological constant is higher than in a
standard cold dark matter Universe and it is precisely because both
the number density and size of large dark matter halos evolve at
different rates in popular cosmological models that observations of
galaxy clusters provide an important discriminator between rival
cosmologies.

The simulations we have carried out follow 2 million gas and 2 million
dark matter particles in a box of side $100\Mpc$.  We have performed
both a SCDM and a $\Lambda$CDM simulation
with the parameters; $\Omega=1.0$, $\Lambda=0.0$, h=0.5,
$\sigma_8=0.6$ for the former and $\Omega=0.3$, $\Lambda=0.7$, h=0.7,
$\sigma_8=0.9$ for the latter.  The baryon fraction was set from Big
Bang nucleosynthesis constraints, $\Omega_bh^2=0.015$ and we have
assumed an unevolving gas metallicity of 0.3 times the solar value.
These parameters produce a gas mass per particle of
$2\times10^9\Msun$.

These simulations \cite{P99a} produce a set of galaxies that fit the
local K-band number counts \cite{G97}. The brightest
cluster galaxies contained within the largest halos are not
excessively luminous for a volume of this size, unlike those found in
previous work \cite{KW93, L99} and presumably \cite{SO98}  
(although they do not state a central
galaxy mass or galaxy luminosity). The fraction of the baryonic material that
cools into galaxies within the virial radius of the large halos in our
simulation is typically around 20 percent, close to the observed baryonic
fraction in cold gas and stars. This is much less than the
unphysically high value of 40 percent reported by \cite{KW93}.

\begin{figure*}
 \centering
\psfig{file=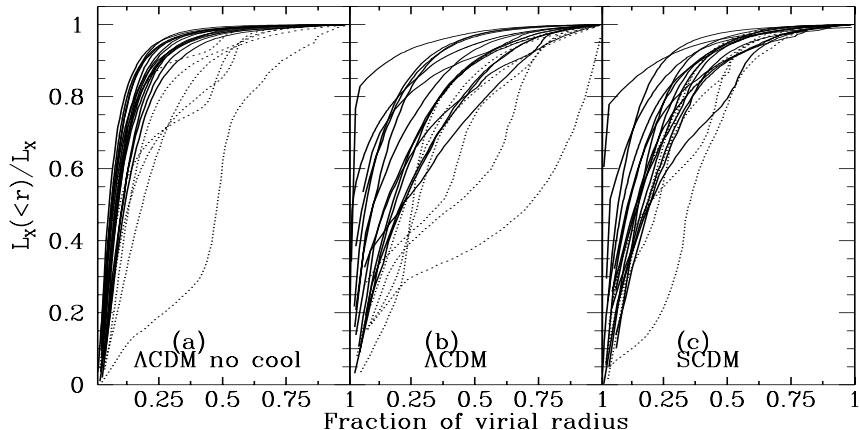,height=6cm,width=12cm}
\caption{The fraction of the total 
bolometric luminosity that is emitted from within the specified radius
for each of the clusters (calculated from equation~1). 
Clusters with significant substructure are shown as dotted lines.
}
\label{fig.lxt}
\end{figure*}

\section{Results}

For each of the 20 largest clusters from each simulation 
we follow \cite{N95} in using the following estimator
for the bolometric X-ray luminosity of a cluster,
\begin{equation}
L_X = 4 \times 10^{32}
\sum \rho_i T_i^{1 \over 2} \thinspace{\rm erg \thinspace s}^{-1}
\label{eq:lx}
\end{equation}
where the sum is over all
the gas particles with temperatures above $12000\K$ 
within the specified radius. Temperatures are in Kelvin and
densities are relative to the mean gas density in the box.
We plot these bolometric luminosities as a function of radius 
for each of our relaxed clusters in figure~1. For the simulation
without cooling the clusters are several times more luminous than
those from the cooling run. This contradicts previous results
\cite{KW93, SO98, L99} 
who all found the X-ray luminosity increased if cooling was turned on.
The cooling clusters are less luminous than 
their counterparts in the simulation without cooling because 
they have lower central temperatures and similar central densities. 
Most of the emission coming from the non-cooling clusters comes
from the central regions, with little subsequent rise in
the bolometric luminosity beyond 0.3 times the virial radius whereas
for the majority of the cooling clusters the bolometric
luminosity continues to rise out to the virial radius.

There has been much debate in the literature centering on the X-ray
cluster $L_X$ versus $T$ correlation. The emission weighted 
mean temperature in keV is
plotted against the bolometric luminosity within the virial
radius for all our clusters in figure~2. The filled symbols
represent the relaxed clusters and the open symbols denote those
clusters that show significant substructure. Clearly the
simulation without cooling produces brighter clusters at the same
temperature. All 3 sets of objects display an $L_X-T$ relation
although there are insufficient numbers to tie the trend down
very tightly. 
Also plotted in figure~2 are the observational data \cite{D95}. 
Our clusters are smaller and cooler because they are not
very massive (due to our relatively small computational volume)
but span a reasonable range of luminosities and temperatures.

\section{Discussion}

Implementing cooling clearly has a dramatic effect on the
X-ray properties of galaxy clusters. Without cooling our
clusters closely resemble those found by previous authors
(\cite{E98} and references therein). These clusters appear
to have remarkably similar radial densities
and bolometric X-ray luminosity profiles, especially when those with
significant substructure are removed.  

With cooling implemented the cluster bolometric X-ray luminosity 
profiles span a broader range.
The formation of a central galaxy within each halo acts to steepen
the dark matter profile, supporting the conclusion of
the lensing studies \cite{K96} that the underlying
potential that forms the lens only has a small core. 
For the largest cluster, a significant amount of baryonic material has
cooled and built up
a large central galaxy. This localised mass deepens the potential well
and contains hot gas with a steeply rising density ($\rho \propto
r^{-2.75}$ in the inner regions). For this cluster around 80 percent
of the bolometric X-ray emission comes from the galactic region and 
this must therefore be viewed as a lower limit as the central emission is
unresolved. Such a large central spike to the X-ray emission 
is already only weakly consistent with the latest observational data \cite{EF99}. For the
remaining 19 clusters the central galaxy is not so dominant and a
shallower central potential well is formed. In these cases the slope
of the central hot gas is $\rho \propto r^{-0.5}$ and the total X-ray emission
is well resolved. 
In principle, the presence of a large galaxy could resolve the
problem of the slope of the X-ray luminosity - temperature relation.
In large clusters, large central galaxies are more likely to be
present and this galaxy deepens the local potential well, boosting 
the emission above the theoretically
expected $L_X \propto T^2$ regression line. Getting a reasonable 
amount of material to cool into the central galaxy is seen to be
of vital importance.

\begin{figure*}
 \centering
\psfig{file=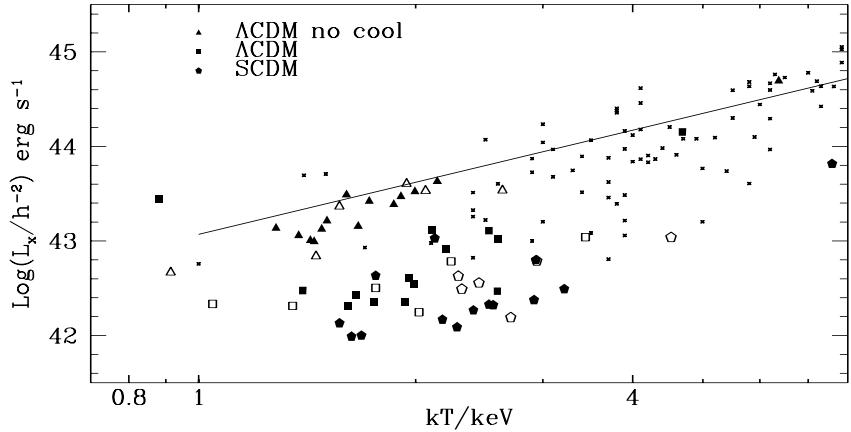,height=6cm,width=12cm}
\caption{The luminosity-temperature relation for all the
clusters. Open symbols refer to those clusters with significant
substructure, filled to those without. The small symbols are the
observation data \cite{D95}. 
}
\label{fig.lxtall}
\end{figure*}

\section{Conclusions}

We have performed two N-body plus hydrodynamics simulations of 
structure formation within a volume of side $100\Mpc$, including
the effects of radiative cooling but neglecting star formation
and feedback. By repeating one of the simulations without radiative
cooling of the gas we can both compare to previous work and
study the changes caused by the cooling in detail. 
A summary of our conclusions follows.

\noindent
(a) The bolometric luminosity for the clusters with radiative cooling
is around five times lower than for matching clusters without it.
Except for the largest cluster where the massive central galaxy
produces a deep potential well the X-ray luminosity profile is less
centrally concentrated than in the non-cooling case with a greater
contribution coming from larger radii. This effect assists in
convergence as we are less dependent upon the very centre of the
cluster profile.

\noindent
(b) The spread of the 
X-ray luminosity -- temperature relation is well reproduced
by our clusters.
Our non-cooling clusters lie close to the regression line suggested
by \cite{E98} and have a similar slope ($\rho \propto r^{-2}$).
We suggest that the increasing
dominance of a large central galaxy on the local potential may produce
the luminosity excess that drives the observed X-ray luminosity -- temperature
relation away from the theoretically predicted slope.

\vfill
\end{document}